\documentclass[letterpaper,american,reprint, aps, prl, superscriptaddress]{revtex4-1}
\usepackage[T1]{fontenc}
\usepackage[latin9]{inputenc}
\usepackage{units}
\usepackage{amsmath}
\usepackage{amssymb}
\usepackage{graphicx}
\usepackage{hyperref}

\begin{document}
\title{SquidLab - a user-friendly program for background subtraction and
fitting of magnetization data}
\author{Matthew J. Coak$^*$}
\affiliation{Department of Physics, University of Warwick, Gibbet Hill Road, Coventry
CV4 7AL, UK}
\affiliation{Cavendish Laboratory, Cambridge University, J.J. Thomson Ave, Cambridge
CB3 0HE, UK}
\affiliation{Center for Correlated Electron Systems, Institute for Basic Science,
Seoul 08826, Republic of Korea}
\affiliation{Department of Physics and Astronomy, Seoul National University, Seoul
08826, Republic of Korea}
\author{Cheng Liu}
\affiliation{Cavendish Laboratory, Cambridge University, J.J. Thomson Ave, Cambridge
CB3 0HE, UK}
\affiliation{CamCool Research Ltd. 44 Thornton Close, Girton, Cambridge, CB3 0NG,
UK}
\author{David M. Jarvis}
\affiliation{Cavendish Laboratory, Cambridge University, J.J. Thomson Ave, Cambridge
CB3 0HE, UK}
\author{Seunghyun Park}
\affiliation{Center for Correlated Electron Systems, Institute for Basic Science,
Seoul 08826, Republic of Korea}
\affiliation{Department of Physics and Astronomy, Seoul National University, Seoul
08826, Republic of Korea}
\author{Matthew J. Cliffe}
\affiliation{School of Chemistry, University of Nottingham, University Park, Nottingham
NG7 2RD, UK}
\author{Paul A. Goddard}
\affiliation{Department of Physics, University of Warwick, Gibbet Hill Road, Coventry
CV4 7AL, UK}
\date{\today}

\begin{abstract}
We present an open-source program free to download for academic use with full user-friendly graphical interface for performing flexible and robust background subtraction and dipole fitting on magnetization data. For magnetic samples with small moment sizes or sample environments with large or asymmetric magnetic backgrounds, it can become necessary to separate background and sample contributions to each measured raw voltage measurement before fitting the dipole signal to extract magnetic moments. Originally designed for use with pressure cells on a Quantum Design MPMS3 SQUID magnetometer, SquidLab is a modular object-oriented platform implemented in Matlab with a range of importers for different widely-available magnetometer systems (including MPMS, MPMS-XL, MPMS-IQuantum, MPMS3 and S700X models), and has been tested with a broad variety of background and signal types. The software allows background subtraction of baseline signals, signal preprocessing, and performing fits to dipole data using Levenberg-Marquadt non-linear least squares, or a singular value decomposition linear algebra algorithm which excels at picking out noisy or weak dipole signals. A plugin system allows users to easily extend the built-in functionality with their own importers, processes or fitting algorithms. SquidLab can be downloaded, under Academic License, from the University of Warwick depository (\href{http://wrap.warwick.ac.uk/129665}{wrap.warwick.ac.uk/129665}).
\end{abstract}
\maketitle

\section*{Introduction}

The SQUID magnetometer \citep{Fagaly2006}, ubiquitous to physics labs worldwide, is an incredibly sensitive instrument capable of measuring magnetic moments down to the absolute quantum limit. SQUIDs are used within cryostats allowing control over temperature and magnetic field to accurately measure the magnetic properties of a huge range of materials and systems, from millikelvin temperatures to well above room temperature and in large magnetic fields. An unavoidable problem however is that the sample must be mounted or supported in some manner - often bulky or complex sample environments are required in the sample region - and the SQUID coils will of course measure both the sample and the `background' from this mechanism. In many cases, the magnetic moment of the sample will be so much larger than this background - which is of course chosen to be as non-magnetic as possible - that the background can simply be disregarded. However, a variety of experiments commonly
push the boundary of this signal/noise ratio: very small or magnetically dilute samples, diamagnetic samples which must be separated from the diamagnetic sample holder and sample environments with large magnetic background contributions such as pressure cells or NMR liquid vials to name a few - see Refs \citep{Kamenev2006,Stamenov2006,Sawicki2011} for a selection of useful discussion. These large, sometimes asymmetric, magnetic background responses can also change and shift position with temperature - making simplistic background modelling and subtraction impossible. The concept of subtracting the background signal prior to dipole fitting, or other data extraction techniques (see e.g. Ref \citep{Cabassi2010}) is well-established, but currently no software tools exist to make this kind of operation accessible.

A SQUID magnetometer works by moving a (magnetic dipole) sample along the $z$ axis of its coaxial superconducting coils, and measuring the induced voltage at various positions along the axis. This then results in a voltage-position curve which is fitted with a dipole form to give a magnetic moment for the datapoint at a fixed temperature and field. This assumes that the recorded signal results indeed from a simple dipole. This does not hold in the case of a significant background, in which the raw voltage signal will not typically be of this form. The raw voltages must instead be subtracted at each position, then the result fitted with the dipole equation to give a background-subtracted magnetic moment $m$.

The Quantum Design Magnetic Property Measurement System (MPMS and MPMS3) models form the most widely adopted commercial SQUID magnetometers. This article will focus mainly on the use of these systems, and SquidLab was primarily designed and tested with these systems, but importers for other systems such as the Cryogenic Inc S700X cryostat are included
(the user can easily implement their own custom importer plugins - most simply by modifying one of the nine included importers), and the operating procedures will be mostly equivalent.

We present in this work our software `SquidLab', an open-source program free to download under an Academic Licence, written and run in Matlab (2019b and above) with both powerful command-line and scripting tools and a full user-friendly graphical user interface (GUI) for performing flexible and robust background subtraction and dipole fitting on magnetization data. Data collected as a function of either temperature $T$ or magnetic field $H$ are supported. A key feature of the design is a plugin system which allows users to easily extend the built-in functionality with their own importers, processes or fitting algorithms, which are then
automatically included into the GUI with rich help and tooltip text. The same Levenberg-Marquadt dipole fitting algorithm used internally in MPMS systems is implemented, as well as a singular value decomposition linear algebra algorithm \citep{BrownThesis2017} which excels at picking out noisy or weak dipole signals. An easy to follow step-by-step GUI is provided to quickly and easily perform background subtraction and fitting operations and to view and export the resulting data,
but a set of command-line and scripting APIs are also provided to allow automated batch processing of large amounts of data. The whole operation of processing a file through the GUI typically takes less than a minute.

While the older generation of MPMS systems included a built-in background subtraction option, the latest MPMS3 instruments have no such functionality and an external solution is required. Even when a built-in option is available, we have repeatedly found that the ability to repeat the background subtraction operation after the experiment has concluded, with a powerful set of options of background datafile, offset compensation and data processing is invaluable in many cases. In addition, the quality of the background subtraction and fits can be improved with a number of methods, which we document here. We have found SquidLab
to be a powerful asset in our own physics research and that of many of our collaborators and believe it will prove likewise to a wide variety of condensed-matter physics and materials chemistry researchers.

\begin{figure*}
\includegraphics[width=2\columnwidth]{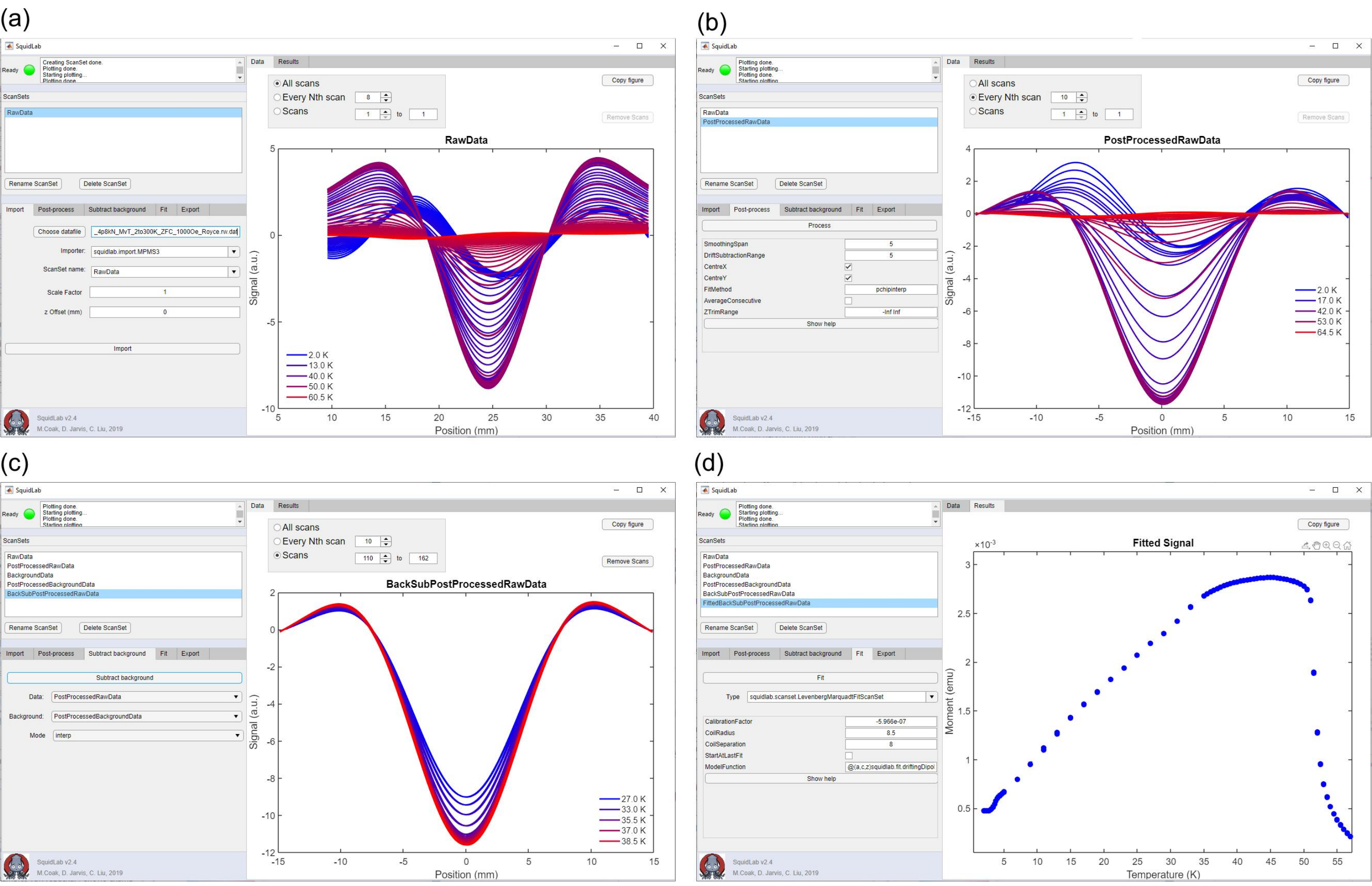}
\caption{\label{fig:SquidLabScreenshots}Screenshots of SquidLab being used
to process a dataset taken on an MPMS3 on 2D ferromagnet VI$_3$ within
a piston-cylinder pressure cell \citep{Son2019}. (a) The raw data
following import (after pressing the Import button shown). (b) The
data after the PostProcessing step. The Post Processing stage performs
several (optional) operations on the raw data, including smoothing,
centering and subtraction of a linear voltage drift. The Import and
Post Processing steps are repeated for a seperate background data
file. (c) The data following background subtraction. (d) The same data
after fitting to the dipole equation, with MPMS3 geometric and scaling
factors shown entered automatically in their relevant boxes. The fits
are shown in the Data tab and the resulting moments plotted as $m$
(emu) vs $T$ in the Results tab when the fit is completed.}
\end{figure*}

\section*{Program Details}

SquidLab is written in Matlab, as both a command-line utility and incorporated into a full GUI using the latest AppDesigner framework to build a robust and platform-agnostic interface with easy access to plotting and graphing tools. It is designed to be fully object-oriented, with abstract classes to allow easy implementation of the many plugin options, particularly the Importer classes. Core to the operation of SquidLab is the ScanSet object, which holds voltage vs position data, accompanying metadata, and any fit results that may have been found. Performing any operation in the GUI, such as `Process' or `Subtract background' creates a new ScanSet of the relevant type to perform the desired operation encoded within it and to hold the result of the operation, with a descriptive name appending this operation, and adds an entry to the list of ScanSets. Clicking through the ScanSets displayed then allows the user to view data at any stage along the data analysis process for comparison and validation, and also to return and repeat steps with alternative options - data is added, not overwritten, by each operation. Multiple data and background ScanSets can be loaded into a single `session', named by the user for ease of identification.

The plugin system works by examining the contents of the relevant namespace directories upon program start and instantiating examples of each class file found in each, which must implement the associated abstract class. Creating a new import plugin, for example, is then as simple as making a copy of an existing import file and editing the relevant code or parameters to describe the desired system. This plugin based design and the availability of all the underlying code mean that users with specific or exotic requirements can customize the code to meet their needs and add new custom functionality.

The selected ScanSet is automatically plotted in the axes of the main GUI tab. After a Fit operation has been carried out, its resulting magnetic moment data will be plotted against field or temperature in the Results tab axes. The ScanSet data shown in the main axes can be selectively visualised through a panel which allows selecting a plotting option. All scans can be plotted, every Nth scan (with N entered) for datasets with a large number of scans, or a range of scans to plot can be selected. This last option allows the user to click through individual scans one at a time, as both upper and lower bound values will increment or decrement together. In this mode, individual scans can be deleted from the ScanSet, to exclude temperatures or fields with jumps or noise in the data or background.

While subtracting one dataset from another is in principle a simple operation, a major obstacle that must be overcome in performing background subtraction is data interpolation. Robust data interpolation methods, handling all edge cases and data types, are a core component of SquidLab. Background and sample scans will not in general be at exactly the same field, temperature and $z$ points - particularly in the case of manually offsetting $z$. Multidimensional interpolant surfaces
must therefore be created from the background scans to allow mapping onto specified sample data points. Constructing these (by making use of Matlab's \emph{scatteredInterpolant} algorithm), robustly with regards to different data files and formats, and dealing with the many edge cases and artefacts is all handled `under the hood' by SquidLab.

\section*{Procedure}

We note that if the magnetometer system used is equipped with a Vibrating Sample Magnetometer option, background subtraction can be very simple and may not require the use of SquidLab. The sample holder and the sample holder plus sample, can simply be measured using the exact same sequence and identical manual positioning/centering. Then, the resulting fitted moments can be subtracted one from the other - (moment{[}sample+background{]}-moment{[}background{]}). This generally works very well, as for the VSM measurement, the sample is oscillating with a small amplitude in the middle of the detection coils and a lock-in amplifier (looking at the 2\textsuperscript{nd} harmonic) then measures the induced voltage, which is in-turn proportional to the magnetic moment via calibrations. Consequently, there is no complicated curve-fitting or waveforms involved, making the procedure very straightforward.

Returning to the normal DC SQUID case and background subtraction using SquidLab, we note that it is always important and easier to minimize the measurement background from the outset. Additionally, if a background is to be subtracted, it is crucial to ensure that the sample+background and the background-only are measured under identical conditions and with identical $z$ positions. An option exists in the SquidLab import stage to compensate for relative offsets in $z$ between the datasets, but this is not recommended as normal procedure. We point out that the 'raw' data files from an MPMS of voltage against position are
required, not just the default extracted moment data files.

While the program is flexible, and operations can be carried out in alternative orders or omitted, the `standard' SquidLab workflow for performing a background subtraction and moment fitting operation would involve four steps - importing the data and background files, post-processing the data and background files, subtracting background from data, and fitting the result. Screenshots of the program being used to carry out these steps are shown in Fig. \ref{fig:SquidLabScreenshots}. The final resulting moments, or the whole ScanSet object, can then be exported to \emph{.txt} file, to the Matlab workspace or plotted in a new figure.

\subsubsection*{Import}

The first step in the SquidLab process is carried out in the Import tab, the leftmost. The user browses for a raw data file to load (for an MPMS or MPMS3 this will be the .raw file generated during measurement) and selects import options and a name for the ScanSet, then clicks the Import button to import the data and create the initial ScanSet. This will then be displayed in the ScanSets list and its data plotted. A drop-down selection list allows selection of an Importer class. This list is automatically populated by loading each plugin file found in the Import directory on program start - this allows a user to quickly and robustly create import functions for any data not handled by the default included plugins.

\subsubsection*{Post-process}

The post-processing step carried out on the Post-process tab allows performing a selection of data processing operations to aid the later subtraction and fitting. A comprehensive built-in Help window is included to guide the user through selecting the appropriate options. The post-processing step is critical to achieving good background subtraction and fitting results.

The data can be smoothed to a selectable degree, which can remove noise in the signal and provide improved fitting robustness. A linear voltage drift is subtracted, using a user-defined number of points at the start and end of each $z$ scan. This ensures that the coefficient of the linear term in a Levenberg-Marquadt fit is close to zero, which improves convergence. The data can optionally be centered in both voltage and $z$; this also allows easier convergence of the dipole fits. These centering operations, and the subtraction of a linear drift, are equivalent to the data processing steps carried out internally in an MPMS system prior to fitting the resulting dipole. The data can additionally be constrained to a selected $z$ range, and an `Average Consecutive' option sets each pair of scans to be averaged for cases (such as the MPMS3) where scans are split into increasing $z$ and decreasing $z$ components.

\subsubsection*{Subtract background}

On the Subtract Background tab the core operation of SquidLab is carried out - a sample and a background ScanSet are selected from drop-down lists of the current ScanSets, and clicking Subtract Background will create the result of the subtraction. Care must be taken with the background subtraction, because the sample and background scans will seldom be at the same set of field or temperature points. Two subtraction modes are provided to be selected from by default, interpolation or nearest-point.

Nearest-point subtraction is the simpler option. This attempts to match every temperature, field and $z$-point in the sample scans to the closest point in the background scans. This approach is easier to follow but will typically give less accurate results, especially if the sample or background scans have widely-spaced temperature or field measurements (in which case the ``nearest point'' may be not very near).

The interpolation background subtraction will usually give superior results. This creates a scattered interpolant object in {[}Position Temperature/Field Signal{]} space, then uses linear interpolation to estimate the background measurement voltage at the sample measurement's position and temperature.

\subsubsection*{Fit}

Fitting options for the background-subtracted data can be specified in the Fit tab, and then selecting the Fit button will fit the dipole (these fits will be overlayed onto the data) and the resulting magnetic moment vs temperature/field plotted in the Results plotting tab. Fitting algorithms are, as with Importers, written as extensible plugins and dynamically loaded at runtime. Two fit options are included as default: `Levenberg-Marquadt' and `SVD' (singular value decomposition), but the user can easily add additional functions for any specific requirements. Fitting and calibration options and scale factors are entered into the relevant input boxes on the Fit tab prior to running the fit. These will be pre-populated with instrument defaults by the Import stage for convenience. The panel of input boxes and controls, and the Help text, are auto-generated for each fit from its public properties and parsed comment text. This means that fit objects with differing input parameters and options can be specified by simply
saving new class files into the `+fit' and `+scanset' folders - the GUI will automatically create appropriate UI controls to display
and edit the properties, and will show any help documentation written into the file.

The fitting algorithms provided by default with SquidLab work as follows. If the `Levenberg-Marquadt' fit option is selected, the processed waveforms of the voltage against position will be fitted to the dipole form \citep{MPMS_app_note_1014-213}:
\begin{alignat}{1}
f(Z) & =X_{1}+X_{2}\cdot Z+X_{3}\cdot\left\{ 2\left[R^{2}+\left(Z+X_{4}\right)^{2}\right]^{-\frac{3}{2}}\right.\nonumber \\
 & -\left[R^{2}+\left(\mathit{\Lambda}+\left(Z+X_{4}\right)\right)^{2}\right]^{-\frac{3}{2}}\nonumber \\
 & \left.-\left[R^{2}+\left(-\mathit{\Lambda}+\left(Z+X_{4}\right)\right)^{2}\right]^{-\frac{3}{2}}\right\} 
\end{alignat}

using a non-linear least squares fitting algorithm, where $f(Z)$ is the SQUID voltage as a function of the sample position $Z$ and
the $X_{i}$ are free parameters for the fit. The constants in this equation are the longitudinal radius, $R$, and the longitudinal SQUID coil separation, $\mathit{\Lambda}$. The fitting parameter $X_{1}$ is a constant offset voltage and $X_{2}$ a linear electronic SQUID drift over time during data collection - these should both be small as they are compensated for during the Post Processing step. The parameter $X_{4}$ is the shift of the sample along the $z$ direction off center. Generally the fits are found to struggle if this strays too far from zero - hence the CentreX option in the Post Processing stage which ensures the data are centered on zero. The parameter $X_{3}$ corresponds to the amplitude, and is used to calculate the magnetic moment of the sample. This can then be multiplied by an instrument-specific calibration factor to obtain the magnetic moment in units of emu-
the desired end result. Default values for this and for geometrical factors $R$ and $\mathit{\Lambda}$ are encoded into the metadata provided by each Importer file and auto-populate the relevant fields in the GUI. For an MPMS3, the default calibration factor (exact value is instrument dependent and can be found with the Pd test sample as the manual instructs) is $(5.966\pm0.293)\times10^{-7}$, $R=8.5$~mm and $\mathit{\Lambda}=8.0$~mm. For an MPMS, the default calibration factor is $(1.096)\times10^{-3}$, $R=9.7$~mm and $\mathit{\Lambda}=15.19$~mm \citep{MPMS_app_note_1014-213} and again can be calibrated. Of course, one can verify that the correct factors are being employed by comparing Squidlab's output for a large-moment sample which does not require background subtraction with the magnetometer's extracted moment values. The user can edit the calibration factor used in the GUI, or specify a new default in the code files for convenience.

Singular value decomposition (SVD) provides an alternative method for fitting the dipole, using a linear-algebra technique which does not rely on estimating start points for a Levenberg-Marquadt fit \citep{Logg2015,S700X-Manual}. This technique does not seem widely used, but it works quite effectively to extract small signals from large backgrounds, and we have found it invaluable for high-pressure magnetometry e.g. in anvil cells \citep{BrownThesis2017}.

SVD considers the measured signal $V_{sig}\left(z\right)$ at a given temperature and field as a superposition of multipole terms:
\[
V_{sig}(z)=\sum_{j=1}^{N}a_{j}f_{j}\left(z\right)
\]
where $f_{j}\left(z\right)$ is the signal from the $j$th multipole term at position $z$, and the $a_{j}$ are the coefficients of each multipole. We find that an $N$ as low as $4$ can often quite accurately reproduce the signal. The coefficient $a_{1}$ corresponds to the dipole signal of interest, and higher coefficients to non-dipole signals arising from e.g. the non-uniform background of the pressure cell. To find the best fit for $a$, we wish to minimise the value:

\[
r=\left|Fa-V_{sig}\right|
\]
where $a$ is a column vector with $N$ elements, and $F$ is an $M\times N$ matrix whose columns correspond to the values of each multipole term. This can be done by solving:
\[
a_{fit}=VS^{-1}U^{T}V_{sig}
\]
where $USV^{T}=F$ forms the singular value decomposition of $F$; the matrices $U,S,V$ can be found quickly and easily e.g. via the $svd$ function in MATLAB. We take the dipole term of the multipole expansion, $f_{1}$, to be eq. (1) with $X_{1},X_{2},X_{4}=0$; this assumption holds when the signal has already been centered around $Z=0$ with linear drift subtracting, which is a motivation for the Post-Processing step in many analyses. Later multipole terms are computed by numerical differentation of preceding terms. This technique is typically much faster than Levenberg-Marquadt (as it is not iterative), does not require guessing starting values for the $X_{i}$, and appears to work better in the case of a small signal on top of a large background, or otherwise poorly-formed dipole shape, where Levenberg-Marquadt may fail to converge. Because the multipole expansion is truncated, it may give a less accurate reproduction of the data or struggle with noise in the signal. Attempting to include too many multipole terms can also provide too many free parameters, given an accurate but unphysical reproduction of the data. For this reason we recommend the use of Levenberg-Marquadt if a clear dipole signal is visible after background subtraction, falling back to SVD in the case of a poorly-formed signal.

\section*{Examples and demonstrations}

\begin{figure*}
\centering{}\includegraphics[width=2\columnwidth]{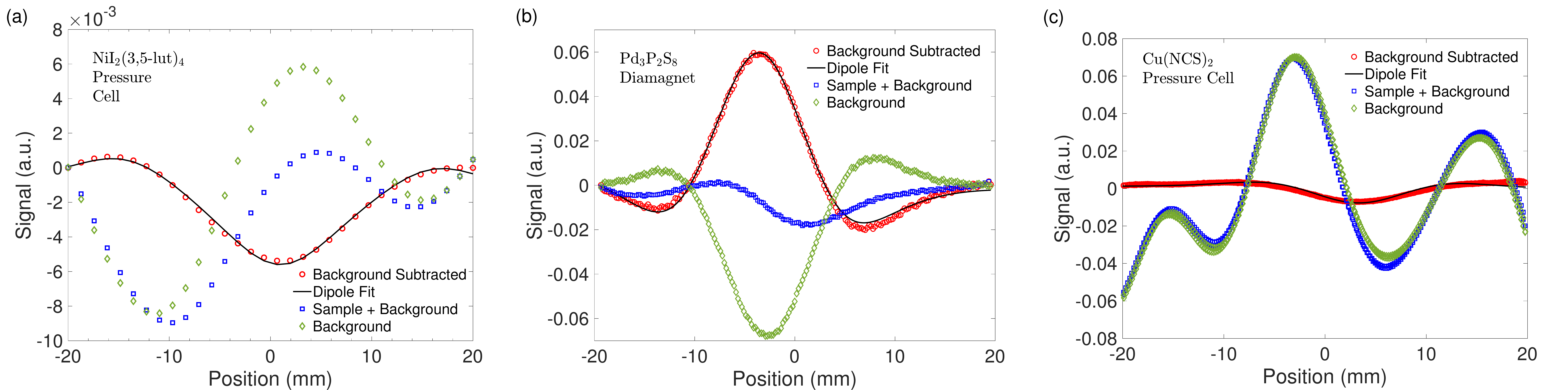}
\caption{\label{fig:Examples-rawVsBackgr}Example dipole signal vs position
plots for a selection of samples and environments, showing the raw
sample+background signal, the background signal and the background-subtracted
result, with the dipole fit carried out in SquidLab which yields the
resulting moment. (a) 1D molecule-based magnet NiI$_2$(3,5-lutidine)$_4$
\citep{Williams2019}, a powder dispersed in the pressure medium of
a piston-cylinder pressure cell, at 1.8~K and 3.6~T. The full $m$
vs $H$ curve was produced as shown in this example - we show only
a single point for clarity. (b) Diamagnetic sample Pd$_3$P$_2$S$_8$
mounted on a standard quartz sample holder, at 5~K and in 100~Oe
fixed field. Without background subtraction, a dipole fit would be
dominated by the (small) paramagnetic moment of the sample holder.
(c) Molecular framework antiferromagnet Cu(NCS)$_2$ \citep{Cliffe2018}
pressed pellet within a pressure cell, at 8~K and 100~Oe. In this
sample the sample moment signal is 1/10th the size of the background,
but is still fully recovered by background subtraction.}
\end{figure*}

In this section we present a selection of example data to demonstrate the power and flexibility of the background subtraction technique using SquidLab, as well as a large-moment sample to show the program reproducing the fitted data generated by the MPMS itself in this case where the background is negligible.

Fig. \ref{fig:Examples-rawVsBackgr} shows examples of typical voltage-position raw scans, with the raw sample+background scan, a background-only scan and the resulting background-subtracted data for each. The dipole fits performed by SquidLab are shown as black lines through the background subtracted data. The example samples shown are (a) an $S=1$ 1D molecule-based
magnet \citep{Williams2019}, with low density of magnetic ions, dispersed in the pressure medium of a piston-cylinder pressure cell (Quantum Design and partners), (b) diamagnetic sample Pd$_3$P$_2$S$_8$ mounted on a standard quartz sample holder and (c) a molecular framework antiferromagnet pressed pellet within a pressure cell (Camcool Research Ltd) \citep{Cliffe2018}.
In all these cases there is no evident dipole in the raw data at the sample position and hence fitting and extraction of the magnetic moment is not possible. Performing the background subtraction however reveals clear near-ideal dipole forms in each case. The case (c) is particularly striking - in this case the sample dipole is an order of magnitude smaller than the pressure cell background, and yet the signal is clearly recovered.

\begin{figure}
\centering{}\includegraphics[width=1\columnwidth]{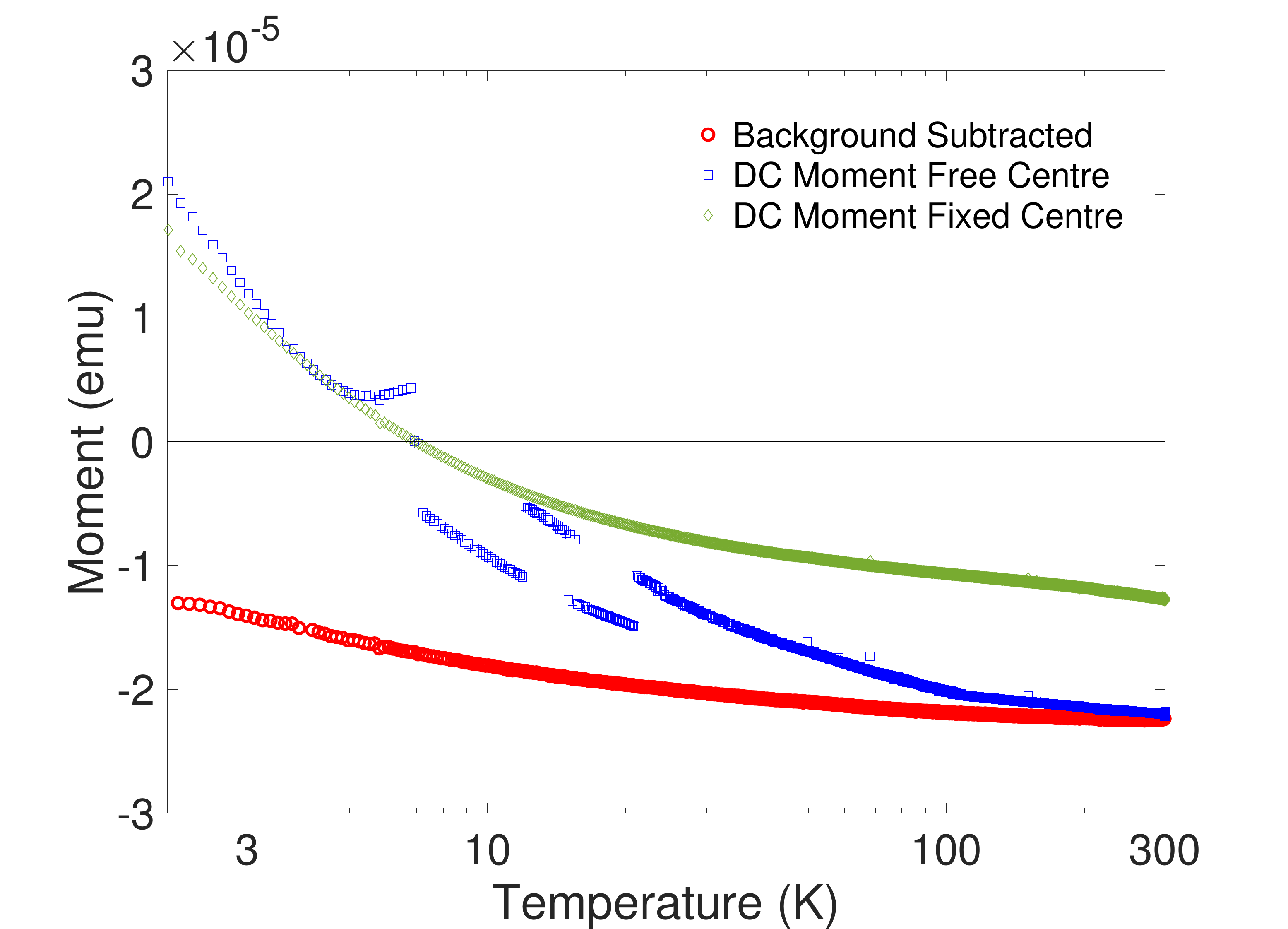}\caption{\label{fig:ExampleDiaSampleMoment}Comparisons of magnetic moment
against temperature taken on an MPMS3 under 100~Oe of field on diamagnetic
sample Pd$_3$P$_2$S$_8$ mounted on a standard quartz sample holder.
The red circles show the resulting moment from performing a background
subtraction within SquidLab - correctly reproducing the negative diamagnetic
moment. The blue squares and green diamonds show the dipole fits returned
by the instrument during the experiment - with a fitted dipole center
and a fixed centre respectively. These give misleading results due
to the presence of the sample holder - in particular, results computed
for the free center case are far from reliable.}
\end{figure}

The resulting magnetic moment vs temperature for the data shown in Fig. \ref{fig:Examples-rawVsBackgr}.b are displayed in Fig. \ref{fig:ExampleDiaSampleMoment} - the background subtracted moment produced by SquidLab, as well as the built-in MPMS3 results with a fixed dipole center and freely fitted dipole center. Neither of the MPMS3 fits are able to correctly describe the negative diamagnetic sample moment, giving extremely misleading results. Treating the data in SquidLab and subtracting the sample holder background, however, correctly produces the expected behavior. A diamagnetic sample such as this is particularly challenging to measure without these background subtraction techniques, as any sample mount will have similar magnetism and be very difficult to separate.

\begin{figure}
\centering{}\includegraphics[width=1\columnwidth]{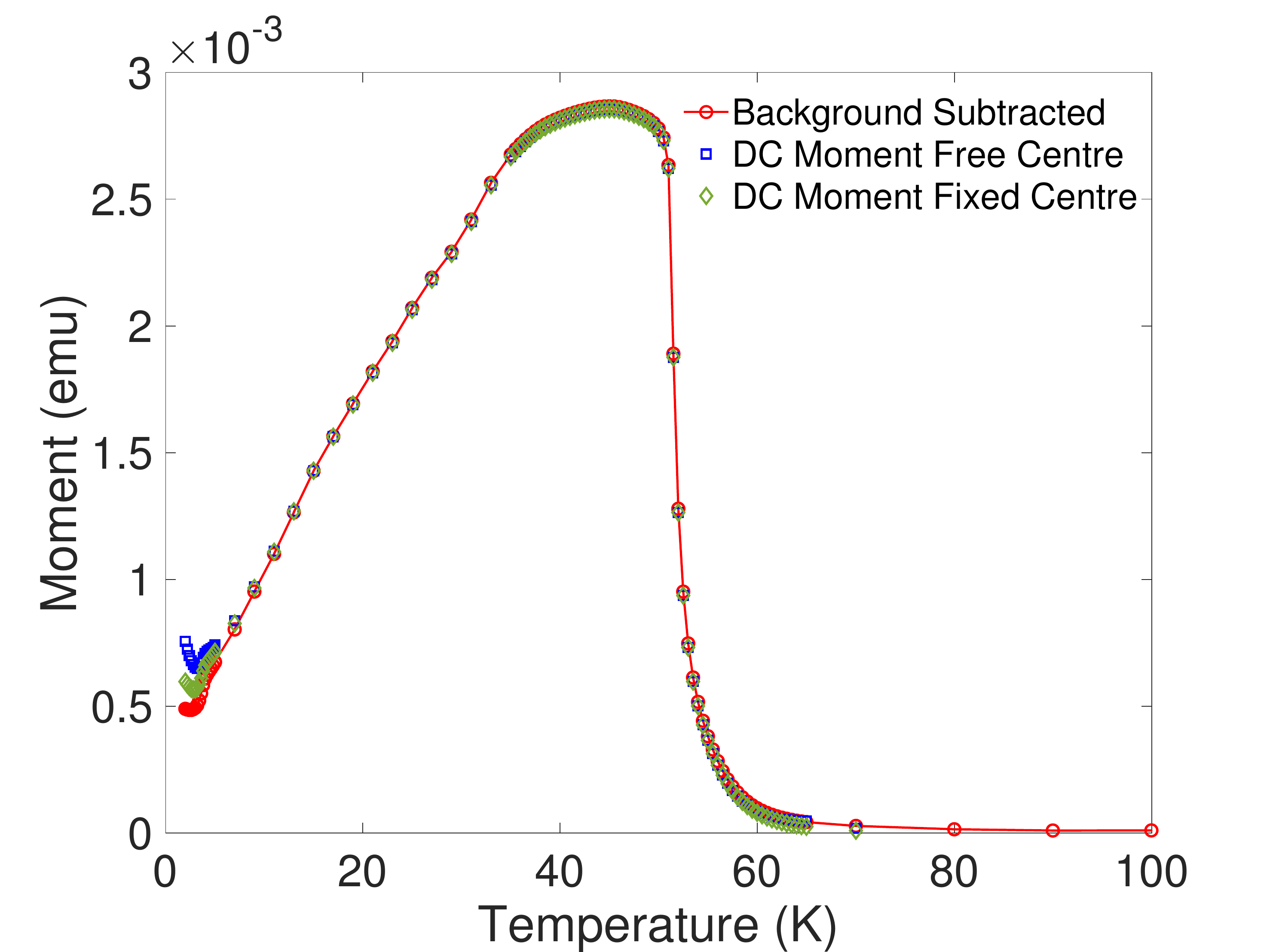}\caption{\label{fig:ExampleVI3FerroMoment}Comparisons of magnetic moment against
temperature taken on an MPMS3 under 1000~Oe of field on large-moment
ferromagnetic sample VI$_3$ \citep{Son2019} measured in a piston-cylinder
pressure cell. The red circles show the resulting moment from performing
a background subtraction within SquidLab; the blue squares and green
diamonds show the dipole fits returned by the instrument during the
experiment - with a fitted dipole center and a fixed centre respectively.
In this case, the sample moment dominates the response and SquidLab
correctly fits and scales the data to match the untreated moments.
At low temperatures however, the paramagnetic response of the cell
becomes significant, leading to an upturn artefact in the unprocessed
data that SquidLab background subtraction removes.}
\end{figure}

To verify the fitting and processing algorithms used, including scaling and calibration factors, we show in Fig. \ref{fig:ExampleVI3FerroMoment} measurements of a ferromagnetic sample with large overall magnetic moment - a moment much larger than that of the pressure cell it is measured in. In this case background subtraction is not truly required, except perhaps at low temperature where the cell has a paramagnetic tail (which the subtraction cleanly removes). This sample is useful
however to verify that our procedures reproduce the internal procedures of an MPMS or MPMS3 system - the fitted dipole with no background subtraction is in this case valid. Very good overlap and agreement is seen between the SquidLab-processed data and the fitted moments from the MPMS. This gives us good confidence that our data handling and fitting procedures are indeed correctly extracting magnetic moments of the correct magnitude from raw voltage data. We have also verified that our background subtraction reproduces the results of an MPMS built-in background subtraction routine where applicable. We have carried out a selection of equivalent tests to verify all the included options.

\section*{Discussion}

We present SquidLab, a free-download open-source program for magnetization background subtraction and fitting written and run in Matlab with a full user-friendly graphical user interface. SquidLab is designed to be flexible and additionally features a plugin system to allow users to extend the built-in functionality. The same Levenberg-Marquadt dipole fitting algorithm used internally in MPMS systems is implemented, as well as a singular value decomposition linear algebra algorithm which excels at picking out noisy or weak dipole signals. SquidLab is a complete solution which covers all steps from data import, handling and processing of data to fitting magnetic moment results to the dipole forms - in a flexible, powerful and extendible framework. An easy to follow step-by-step GUI is provided to quickly and easily perform background subtraction and fitting operations and to view and export the resulting data, but a set of command-line and scripting APIs are also provided to allow automated batch processing of large amounts of data.

We have shown a selection of examples of background subtraction in action, to show that it is necessary in many cases to avoid completely spurious data, and demonstrated that clear signals can be recovered even from a background ten times the size of the sample dipole. While the original use case was for high-pressure measurements, the technique has proved suitable for a wide variety of samples and environments. We additionally have tested and verified large-moment samples, where the magnetometer can itself fit without the background subtraction, to show that our technique reproduces the same results in these cases.

\begin{acknowledgments}
The authors would like to thank: Samuel Holt, Robert Williams, Samuel Curley and Nahyun Lee for their efforts in measurement tests and feedback on the application, Randy Dumas and Jordan Thompson of Quantum Design for providing useful technical information, Danielle Villa and Jamie Manson for growing samples and Si\^an Dutton, Siddharth Saxena and Je-Geun Park for their generous help, support and discussions. We acknowledge the support of the EPSRC and of Jesus College Cambridge.
This work was supported by the Institute for Basic Science (IBS) in Korea (Grant No. IBS-R009-G1). This project has received funding from the European Research Council (ERC) under the European Union's Horizon 2020 research and innovation programme (grant agreement No. 681260). Data presented in this paper resulting from the UK effort will be made available at https://wrap.warwick.ac.uk/132173.
\end{acknowledgments}

\end{document}